\documentclass{emulateapj}
\usepackage[usenames,dvipsnames]{color}          

\shorttitle{Alfv{\'e}n waves in simulations of solar photospheric vortices}
\shortauthors{S.~Shelyag, P.S.~Cally, A.~Reid, M.~Mathioudakis}

\begin{document}
\title{Alfv{\'e}n waves in simulations of solar photospheric vortices}

\author{S.~Shelyag, P.S.~Cally}
\affil{Monash Centre for Astrophysics, School of Mathematical Sciences, Monash University, Victoria 3800, Australia}
\and
\author{A.~Reid, M.~Mathioudakis}
\affil{Astrophysics Research Centre, School of Mathematics and Physics, Queen's University Belfast, Belfast BT7 1NN, Northern Ireland, UK}

\date{01.01.01/01.01.01}

\begin{abstract}
Using advanced numerical magneto-hydrodynamic simulations of the magnetised solar photosphere, including non-grey radiative transport and 
a non-ideal equation of state, we analyse plasma motions in photospheric magnetic vortices. We demonstrate that apparent vortex-like motions 
in photospheric magnetic field concentrations do not exhibit ``tornado''-like behaviour or a ``bath-tub'' effect. While at each time instance the 
velocity field lines in the upper layers of the solar photosphere show swirls, the test particles moving with the time-dependent velocity field do 
not demonstrate such structures. Instead, they move in a wave-like fashion with rapidly changing and oscillating velocity field, determined mainly 
by magnetic tension in the magnetised intergranular downflows. Using time-distance diagrams, we identify horizontal motions in the magnetic 
flux tubes as torsional Alfv\'en perturbations propagating along the nearly vertical magnetic field lines with local Alfv\'en speed.
\end{abstract}


\section{Introduction}
Vortices in magnetic field concentrations in the simulated upper solar photosphere, first identified by \citet{voegler2, voegler1}, have recently attracted a 
lot of interest. These structures have been suggested as a primary candidate mechanism for energy transport from the solar interior to the outer 
layers of the solar atmosphere \citep{wedemeyer2}. Recently, \cite{shelyag2011} demonstrated that the major contributor to vertical vorticity in 
these photospheric magnetic field concentrations is magnetic tension in the low plasma $\beta$ regions. Initial analysis of velocity fields in 
magnetic photospheric vortices demonstrated a complex multi-layered structure of oppositely rotating surfaces with nearly constant vorticity 
that are parallel to the magnetic field lines \citep{shelyagangeo,fedunangeo}. \citet{steiner2} found a similar behaviour of the velocity 
field in the solar photospheric simulations carried out with CO5BOLD \citep{freytag1}. Poynting flux emerging from the simulated solar photosphere was 
linked to the horizontal plasma motions in the photospheric magnetic field \citep{shelyag2012}. \citet{moll1} showed that vortex motions in the 
simulated photosphere are able to dissipate their energy and produce additional local heating of the solar chromosphere. \citet{kitiashvili1} 
suggested that small-scale solar atmospheric eruptions, such as spicules, are linked to horizontal vortex-like motions in photospheric magnetic 
flux tubes. \citet{morton1} observationally demonstrated excitation of incompressible waves in the solar chromosphere by photospheric vortex 
motions. Using multi-wavelength observations, it was recently demonstrated that these essentially magnetic structures extend from the lower 
photosphere to the corona, and are connected to the observed swirls in the solar chromosphere \citep{wedemeyer1} and lower coronal regions 
\citep{wedemeyer2}.

In this Letter, using a time series of magnetised photospheric models produced by the MURaM code \citep{voegler1}, we further analyse 
the character of plasma motions in intergranular magnetic field concentrations. We demonstrate that, if dependence of the local velocity 
field on time is taken into account, the apparent swirling motions in the strong magnetic field disappear, and swirls do
not exist in photospheric magnetic flux tubes. Instead, a wave-like behaviour is observed for horizontal 
motions of test particles following the time-dependent velocity field. The restoring force of magnetic tension acts as a conduit of 
oscillatory vorticity upward from deeper photospheric layers, where it is generated by convective motions. Using vertical-component-of-vorticity 
time-distance diagrams taken along a vertical axis we show that the vorticity perturbations propagate upward along the nearly-vertical field lines 
of photospheric magnetic field concentrations with time-dependent local Alfv{\'e}n speed, and thus can be identified as torsional Alfv\'en waves. 
This result also explains quasiperiodic oscillations and the short lifetimes of photospheric magnetic vortices, as noted by \citet{moll2} and 
\citet{kitiashvili1}.

This Letter is structured as follows. In Section 2 we describe the simulation setup and the data we use for our analysis of photospheric vortices. 
The time-distance analysis of plasma motions in the upper-photospheric magnetic field concentrations is described in Section 3. Section 4 
concludes our findings.

\section{Simulation setup}

\begin{figure}
\plotone{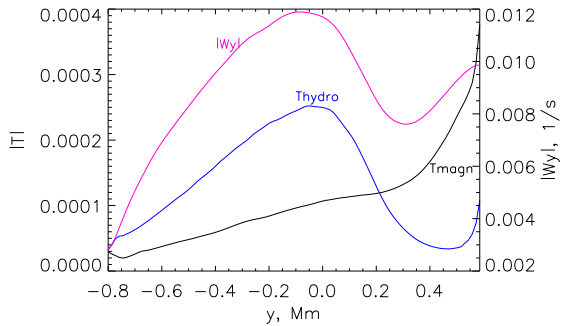}
\caption{Dependences of the moduli of the vertical component of vorticity ($|W_y|$) and of magnetic ($T_\mathrm{magn}$) and hydrodynamic 
($T_\mathrm{hydro}$) vorticity sources on height in Mm in the simulated solar photosphere. Here, negative values of $y$ correspond to the
convection zone, while positive values refer to the photosphere.}
\label{fig1}
\end{figure}

The MURaM radiative MHD code \citep{voegler1} has been used to carry out simulations of the magnetised solar photosphere. The code is well-tested, has 
demonstrated similar performance to other radiative MHD codes available in the community\citep{beeck1}, and has been widely used to 
provide an insight into the physical processes behind various solar observational phenomena, such as, for example, sunspots \citep{rempel1}, magnetic 
flux emergence \citep{cheung1}, and photospheric bright points \citep{shelyagbp1, shelyagbp2}. The code reproduces thermal and radiative properties 
of the solar photosphere well, as demonstrated by comparison of simulated photospheric spectral lines with observations \citep{shelyag2}.

For the analysis presented in this Letter we use the code setup described in detail by \citet{shelyag2011}. The physical domain size is $12 \times 12~\mathrm{Mm^2}$ 
in the horizontal directions and $1.4~\mathrm{Mm}$ in the vertical direction, and the resolution is $25\times 25~\mathrm{km^2}$ and $14~\mathrm{km}$, 
respectively. To generate the photospheric data, we introduce a uniform 200 G vertical magnetic field into a well-developed non-magnetic photospheric 
convection model, and, after the initial transient phase (about 40 minutes of physical time), we record 240 snapshots of photospheric model data with a 
mean cadence of 2.41 seconds. This short time step is required to capture processes at Alfv\'enic time scales in strongly magnetised regions of the solar 
photosphere where the characteristic speed can reach 50--100 $\mathrm{km\,s^{-1}}$. 

During the initial simulation phase, the magnetic field is advected into the intergranular lanes by convective motions, and photospheric magnetic flux 
concentrations with the strength of about 1.7 kG at the continuum formation height and of about 800 G in the upper photosphere are formed. 
Simultaneously, magnetic vortex structures appear in the low plasma $\beta$ regions of the simulated solar atmosphere. 

Figure \ref{fig1} displays the dependences of absolute values of magnetic (black curve) and hydrodynamic (blue curve) sources in the vertical component of 
vorticity according to Equation (4) of \citet{shelyag2011}, as well as the modulus of the vertical component of vorticity itself (pink curve), all averaged horizontally 
over the numerical domain. The figure shows that in the lower photosphere (below 0.2 Mm) the main source of vorticity is compressible fluid motion. 
Conversely, in the strongly magnetised upper photosphere (above 0.2 Mm), magnetic sources dominate over purely hydrodynamic vorticity sources. 
This clearly indicates that magnetic field effects are responsible for a strong increase in vertical vorticity above 0.3 Mm in the photosphere. 
\citet{shelyag2011} demonstrated that the vorticity source term, which contains magnetic tension, is responsible for this effect.

\begin{figure}
\plotone{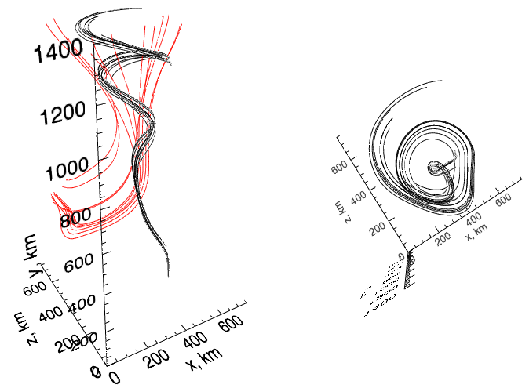}
\caption{Probe particle tracks in the simulated photospheric magnetic field concentration under the action of the velocity field 
obtained from the first snapshot of the data series. Views from the side and from the top of the vortex structure are shown in left and
right sides of the figure, respectively. Also, the magnetic field lines are plotted in grey (online version - red) in the left side of the figure. A
fly-around animation of the left part of the figure is included in online material.}
\label{fig2}
\end{figure}

To visualise the three-dimensional velocity field in as simple and clear a manner as possible, we use a small number of seed particles that are
randomly magnetic field-weight distributed around a magnetic field concentration over a horizontal plane close to the top boundary of the domain. 
This biases them toward the strong-field regions near the tube axis. 

\begin{figure}
\plotone{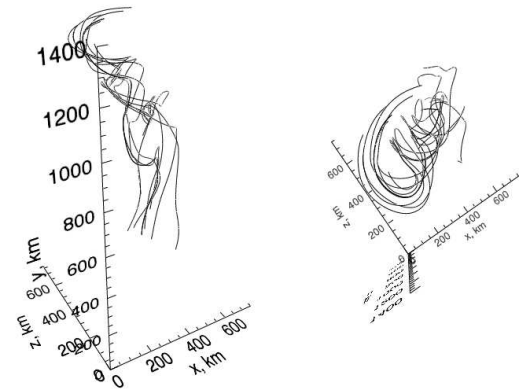}
\caption{Probe particle tracks in the simulated photospheric magnetic field concentration under the action of the time-dependent velocity field 
over the simulation time series (about 500 seconds). A fly-around animation of the left part of the figure is included in online material.}
\label{fig3}
\end{figure}

\begin{figure*}
\plotone{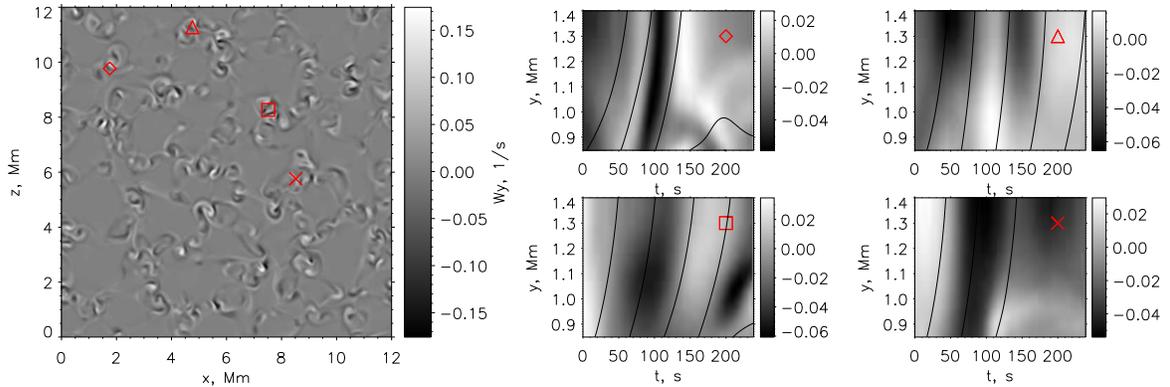}
\caption{Upper photospheric vertical vorticity over a horizontal ($x$-$z$) plane at height about $y=500$ km (left panel) and examples of time-distance 
diagrams of Alfv{\'e}n wave propagation in the upper solar 
photosphere (right panels). The time-distance diagrams are measured along the vertical direction in the simulated solar photosphere at the positions 
marked by the symbols in the left panel. The curves plotted over the time-distance diagrams represent tracks for test particles under the action of
local time-dependent Alfv{\'e}n velocity. They are also corrected to account for the local time-dependent flow field.}
\label{fig4}
\end{figure*}

We first plot probe particle tracks under the action of the velocity field obtained from the first snapshot of the data series. 
An example of probe particle tracks under the action of this velocity field is shown in Fig.~\ref{fig2}. These are streamlines. As is evident 
from the figure, a vortex-like structure is present in the upper part of the simulated photosphere. However, if the probe 
particles with the same initial coordinates are let to evolve in the time-dependent velocity field from the whole time series of the simulation, 
the upper-photospheric vortex completely disappears and is replaced by nearly random, quasi-oscillatory particle tracks (see Fig.~\ref{fig3}). 
These are pathlines. Streamlines and pathlines coincide only in steady flows. Their disparity clearly demonstrates that three-dimensional flow 
in upper-photospheric vortices in regions of strong magnetic field is far from steady. 
\citet{moll1} recently pointed out that the low twist of the magnetic field lines in their modelling indicates a non-stationary character
of the flow. Indeed, the time scale for flow change is similar to the 
flow time scale itself. Thus, processes in the upper-photospheric magnetic vortices cannot be physically related to those in 
sinks or tornadoes in Earth's atmosphere, where steady velocity fields are present. 
Vortical flow in a low-beta atmosphere cannot be sustained over many Alfv{\'e}n crossing times unless it is an essentially solid-body rotation 
induced by rotational flows or flux tube unwindings in the high-beta layers below, since dominant magnetic tension will resist it.

In the following section, we confirm this finding with more detailed analysis of propagation of the vorticity perturbation through the solar atmosphere.

\section{Time-distance analysis of vorticity in photospheric magnetic field concentrations}
Basic analysis of the probe particle motions given in the previous section shows an absence of steady vortex flows in photospheric magnetic field 
concentrations. The presence of both vorticity signs, earlier reported by \citet{shelyagangeo} and \citet{steiner2}, manifests both clockwise and anti-clockwise torsional motions 
within the same magnetic structure. This fact suggests wave-like motions of the plasma in photospheric magnetic field concentrations. To study these 
waves and to identify their types we perform time-distance analysis of the vorticity perturbation along the magnetic field lines in the simulated domain. 
Construction of time-dependent three-dimensional magnetic surfaces is a computationally difficult task, so we choose points where the magnetic 
field is nearly vertical in the neighbourhood of the upper boundary of the domain. 

Figure~\ref{fig4} shows a map of the vertical ($y$-)component of vorticity measured at about 500 km above the average continuum formation 
level (left panel), and four examples of time-distance diagrams of vertical propagation of the vertical component of vorticity. In choosing the examples, 
we restricted the selection to waves that propagate upwards only, though both upward and downward directions of wave propagation are observed. 
As is evident from the plots, all examples show oscillatory patterns of fast-propagating perturbations along the vertical direction. We were able to identify 
the wave type by overplotting the path for a test particle, which moves with the local, time-dependent Alfv{\'e}n speed, according to the equation $\Delta y=
\Delta t \left(\sqrt{B^2 /4\pi\rho} +v_{y} \right)$ in {\it cgs} units, where $v_y$ is local time-dependent vertical flow speed.
The test particle tracks are plotted over the time-distance diagrams and clearly coincide with the oscillatory pattern 
of the vertical component of the vorticity. Notably, the mean vertical flow speed averaged over the time series never exceeds $3~\mathrm{km/s}$, and 
the maximum vertical flow speed is about $6~\mathrm{km/s}$. Therefore, motions of the test particles with speed about $20~\mathrm{km/s}$ cannot be 
explained by vertical plasma flows, as is 
evident from Fig.~\ref{fig4}. This finding demonstrates that the propagation speed for the vorticity perturbation is equal to the Alfv\'en 
speed. In the plots shown in the right part of Fig.~\ref{fig4}, the maximum magnetic field inclination angle in the time-distance region where a clear wave 
propagation pattern is observed is less than $10^\circ$. Therefore, the assumption about equivalence of propagation along the magnetic field line and along the 
vertical direction is well-justified. Summarising the findings, the perturbation in the vertical component of the vorticity, that propagates vertically along magnetic
field lines, is a horizontal torsional velocity perturbation, and thus it can be identified as a torsional Alfv\'en wave. 

In fact, nearly all vertical columns in the regions of strong magnetic field in the simulation domain show propagation of a perturbation with speed close to the Alfv\'en speed. 
Abundance of these patterns in the domain and nearly perfect agreement between the perturbation phase speed and the Alfv\'en speed also rule out a possibility of advection 
of pre-existing vorticity patterns by horizontal bulk flows. Thus, it can be concluded that torsional Alfv\'en waves originating in the higher photosphere are abundant in plage 
regions of the solar photosphere.

\section{Conclusions}
In this Letter, we analysed apparent photospheric magnetic vortex structures in a dynamic realistic simulation of solar photospheric 
magnetoconvection. We identified that no long-lived vortex structures in photospheric magnetic field concentrations exist. Instead, the 
time-dependent velocity field obtained from the simulations shows oscillatory behaviour in the strong intergranular magnetic field regions. Using 
time-distance analysis of the vertical component of vorticity we found that the perturbations propagate vertically along the magnetic field lines 
with speeds equal to the local Alfv{\'e}n speed. Thus we identified the perturbations as torsional Alfv{\'e}n waves in intergranular magnetic flux 
concentrations. In the structures shown in our simulations the velocity pattern changes within approximately 50 seconds to the opposite 
direction. This timescale is considerably shorter than the lifetime of the inter granular magnetic field concentrations which continue to exist 
throughout this process.

As our study shows, we must exercise caution when visualising three-dimensional motions in the turbulent and magnetised solar atmosphere. 
Steady velocity fields cannot be used for analysis, since the speed with which the velocity pattern changes is of the same order as the actual velocity, 
while studies of time-dependent velocity fields impose additional requirements on the simulated data series, such as spatial resolution and cadence,
as well as on the computational resources needed to perform the computations. 

The vorticity perturbations we observe in the simulations, unlike pure Alfv\'en waves, cause perturbations in thermodynamic parameters of the solar
plasma. These perturbations are possibly caused by some unavoidable amount of artificial diffusivity and resistivity included in the code for numerical
stability. Another reason for these perturbations may be that other, magneto-acoustic oscillatory modes are present in the simulated magnetic field
concentrations.

It also should be mentioned that, apparently, photospheric magnetic bright point motions, which originate at the base of the photosphere, do not 
represent good candidates for this type of wave observation, since they are formed deeper in photospheric layers \citep{shelyagbp2,carlsson1}. 
Their measured paths \citep{bonet1} suggest a hydrodynamic, non-oscillatory type of photospheric horizontal motion.

Previous observational and theoretical reports on Alfv\'en waves in the solar chromosphere \citep[e.g.][]{jess1} and in sunspots \citep{brown1, khomenko1} 
clearly suggest that Alfv{\'e}n waves are a common phenomenon in strong solar magnetic fields. Interestingly, recent high-cadence observations of 
prominences show that their tornado-like appearance is ``an illusion due to projection effects'' \citep{panasenco1}, suggesting a general presence of Alfv\'en-type
motions in the solar atmosphere and corona.

Observational confirmation of the result given 
in this Letter is currently rather difficult since it requires spectroscopic and spectropolarimetric observations with both very high resolution of the order of 
20--50~km and extremely high cadence of about 2 seconds. These may become possible with upcoming high-aperture instruments, such as ATST. Further 
investigation of the reaction of spectral line profiles to photospheric Alfv\'en waves will be needed before such an observational attempt.

Another interesting problem is detailed identification of the lower-photospheric sources of 
upper-photospheric torsional waves. Among the possibilities are hydrodynamic vortices in the
deep photosphere, small-scale granular motions and short-lived turbulent eddies. Further theoretical 
and computational work is needed to fully determine the sources of Alfv\'en waves observed in the 
upper photosphere.

\acknowledgments
\section{Acknowledgement}
This research was undertaken with the assistance of resources provided at the NCI National Facility systems at the Australian National University through the 
National Computational Merit Allocation Scheme supported by the Australian Government, and at the Multi-modal Australian ScienceS Imaging and Visualisation 
Environment (MASSIVE) (www.massive.org.au). The authors also thank Centre for Astrophysics \& Supercomputing of Swinburne University of Technology (Australia) 
for the computational resources provided. Dr Shelyag is the recipient of an Australian Research Council's Future Fellowship (project number FT120100057).

\end{document}